\documentclass[pdftex,smallextended,final,epjc3]{svjour3}
\usepackage[a4paper,top=4cm,bottom=4cm,left=2.5cm,right=2.5cm]{geometry}

\RequirePackage{graphicx}
\RequirePackage{kantlipsum,widetext}
\RequirePackage{amsmath}
\RequirePackage{enumerate} 

\usepackage{widetext}
\usepackage[dvipsnames]{xcolor}
\usepackage{physunits}
\usepackage{paralist}

\newcommand{\thetaCO}{\theta}

\RequirePackage{mathptmx}      

\usepackage{newtxtext,newtxmath}

\RequirePackage{latexsym}
\RequirePackage[numbers,sort&compress]{natbib}
\RequirePackage[colorlinks,citecolor=blue,urlcolor=blue,linkcolor=blue]{hyperref}
\journalname{}

\begin{document}

\title{\boldmath Shedding light on the $ X(3930) $ and $ X(3960) $ states with the $B^- \to K^- J/\psi \omega $ reaction}


\author{%
L.~M.~Abreu\thanksref{e1,addr1,addr2}%
\and\nolinebreak
M.~Albaladejo\thanksref{e2,addr2}%
\and\nolinebreak%
A.~Feijoo\thanksref{e3,addr2}%
\and\nolinebreak%
E.~Oset\thanksref{e4,addr3,addr4}%
\and\nolinebreak%
J.~Nieves\thanksref{e5,addr2}%
}

\thankstext{e1}{\href{mailto:luciano.abreu@ufba.br}{luciano.abreu@ufba.br}}

\thankstext{e2}{\href{mailto:Miguel.Albaladejo@ific.uv.es}{Miguel.Albaladejo@ific.uv.es}}

\thankstext{e3}{\href{mailto:edfeijoo@ific.uv.es}{edfeijoo@ific.uv.es}}

\thankstext{e4}{\href{mailto:oset@ific.uv.es}{oset@ific.uv.es}}

\thankstext{e5}{\href{mailto:Juan.M.Nieves@ific.uv.es}{Juan.M.Nieves@ific.uv.es}}


\institute{\setlength{\parindent}{0pt} 
Instituto de F\'{\i}sica, Universidade Federal da Bahia, Campus Universit\'{a}rio de Ondina, 40170-115, Bahia, Brazil \label{addr1}%
\and%
Instituto de F\'isica Corpuscular, Centro Mixto Universidad de Valencia-CSIC, Institutos de Investigaci\'on de Paterna, Aptdo. 22085, 46071 Valencia, Spain  \label{addr2}%
\and%
Departamento de F\'{\i}sica Te\'orica and IFIC, Centro Mixto Universidad de Valencia-CSIC, Institutos de Investigaci\'on de Paterna, Aptdo. 22085, 46071 Valencia, Spain  \label{addr3}%
\and%
Department of Physics, Guangxi Normal University, Guilin 541004, China \label{addr4}%
}

\date{}

\maketitle

\setlength{\parindent}{7pt} 
\begin{abstract}
We have studied the contribution of the state $X(3930)$, coming from the interaction of the $D \overline{D}$ and $D^{+}_s D^{-}_s$ channels, to the $B^- \to K^- J/\psi \omega $ decay. The purpose of this work is to offer a complementary tool to see if the $X(3930)$ state observed in the $D^+ D^-$ channel is the same or not as the $X(3960)$ resonance claimed by the LHCb collaboration from a peak in the $D^{+}_s D^{-}_s$ mass distribution around threshold. We present results for what we expect in the  $J/\psi \omega $ mass distribution in the $B^- \to K^- J/\psi \omega $ decay and conclude that a clear signal should be seen around $3930\,\MeV$. At the same time, finding no extra resonance signal at $3960\,\MeV$ would be a clear indication that there is not a new state at  $3960\,\MeV$, supporting the hypothesis that 
the near-threshold peaking structure peak  in the $D^{+}_s D^{-}_s$ mass distribution is only a manifestation of a resonance below  threshold. 

\end{abstract}

\section{Introduction}\label{sec:intro}
In Refs.~\cite{LHCb:2022vsv,LHCb:2022dvn} the LHCb collaboration analyzed  the $B^+ \rightarrow D_s^+ D_s^- K^+$ decay and reported a peak in the $D^+_sD^-_s$ mass distribution close to the threshold, which was associated  to a new resonance, $X(3960)$. Soon after disclosing the experimental results in Ref.~\cite{LHCb:2022NewObservations}, it was suggested that this peak would be a necessary consequence of a weakly bound  $D_s^+ D_s^-$ state and there was no need to invoke a new resonance if such a state existed \cite{Bayar:2022dqa,Ji:2022uie}. Actually such a state with scalar quantum numbers is predicted in the lattice QCD (LQCD) calculation of Ref.~\cite{Prelovsek:2020eiw}. Indeed, a state of molecular nature coupling to $D_s^+ D_s^-$ and  $D\overline{D}$ is found close to  $D_s^+ D_s^-$ threshold, with a strong coupling to $D_s^+ D_s^-$ and a weaker one to $D\overline{D}$. At the same time another state coupling to $D\overline{D}$ below its threshold is also found in Ref.~\cite{Prelovsek:2020eiw}. Such a  $D\overline{D}$ bound state had been previously predicted in  Ref.~\cite{Gamermann:2006nm,Nieves:2012tt,Hidalgo-Duque:2012rqv}. Although no $D_s^+ D_s^-$ bound state was obtained in Ref.~\cite{Gamermann:2006nm}, its existence with a mass in the $3915 - 3935\,\MeV$ range was predicted in the single-channel calculation carried out in Ref.~\cite{Hidalgo-Duque:2012rqv}. The coupled-channels scheme of Ref.~\cite{Gamermann:2006nm} was reviewed in Ref.~\cite{Bayar:2022dqa}, where it was shown that a small decrease of the $D\overline{D} \to D_s^+ D_s^-$ transition potential reverted into the appearance of a  $D_s^+ D_s^-$ bound state close to the $D_s^+ D_s^-$ threshold. It was explained there that this decrease actually was naturally obtained by considering the full propagator $(q^2-m_{K^*}^2)^{-1}$ in the $K^*$ exchange instead of the $q^2 \to 0$ approximation used in Ref.\,\cite{Gamermann:2006nm}. The results of Ref.\,\cite{Bayar:2022dqa} agree then with those of Ref.\,\cite{Prelovsek:2020eiw}, and a bound state in the $D_s^+ D_s^-$--$\overline{D}D$ coupled-channel amplitude is obtained, coupling more strongly to $D_s^+ D_s^-$ than to $\overline{D} D$, and hence it approximately qualifies as a $D_s^+ D_s^-$ bound state. However, given the fact that this state also couples to $\overline{D} D$, it would be expected to be visible in the $D^+ D^-$ spectrum of some decay. Actually, the LHCb collaboration also finds a scalar state in the $D^+ D^-$ spectrum in the $B^+ \rightarrow D^+ D^- K^+$ decay \cite{LHCb:2020pxc,LHCb:2020bls} branded as $X_0(3930)$ with $J^{PC}=0^{++}$ and with mass and width given by:
\begin{equation}
M=3924\pm2~\rm MeV,\, \Gamma=17\pm5~\rm MeV.
\nonumber 
\label{eq:Xi_1690}
\end{equation}
The $D_s^+ D_s^-$ threshold is at $3936.7\,\MeV$, and hence this state is barely $10\,\MeV$ below the $D_s^+ D_s^-$ threshold, so it could as well correspond to the $D_s^+ D_s^-$ bound state found in Refs.~\cite{Prelovsek:2020eiw,Bayar:2022dqa,Hidalgo-Duque:2012rqv,Ji:2022uie}. This is the suggestion made in Refs.~\cite{Bayar:2022dqa,Ji:2022uie}, where, in addition, it was shown that the existence of this state produces an enhancement above the  $D_s^+ D_s^-$ threshold in the mass distribution in the $B^+ \rightarrow D_s^+ D_s^- K^+$ decay, compatible with the experimental results. In Ref.~\cite{Bayar:2022dqa} it was also reproduced the experimental ratio of the $X \to D^+ D^-$ to $X \to D_s^+ D_s^-$ decay widths, of the order of $0.3$, assuming $X$ to be the same state responsible for the two peaks. In turn, the authors of Ref.~\cite{Ji:2022uie} employed an effective field theory based on Heavy Quark Spin Symmetry (HQSS) to conclude that the bump in the $D_s^+ D_s^-$ mass distribution could be described by a bound or virtual state below threshold. Using the pole (either virtual or bound)  position and the existing information on the $X(3872)$ and $Z_c(3900)$ resonances, a complete spectrum of the $S$-wave hadronic molecules formed by a pair of ground state charmed and anticharmed mesons was established in that work.
In a following paper \cite{Ji:2022vdj} the available data in $D \overline{D}$ and $D_s^+ D_s^-$ channels from both $B$ decays and $\gamma\gamma$ fusion reaction are analyzed, firmly concluding that the $X(3930)$ and $X(3960)$ should correspond to the same state. 

The molecular picture to explain the $D_s^+ D_s^-$ bump is also supported using QCD sum rules in  Refs.~\cite{Xin:2022bzt,Mutuk:2022ckn}. The boson exchange model is used in Ref.~\cite{Chen:2022dad} to obtain also a molecular state with the $D_s^+ D_s^-$, $\overline{D}D$, $D_s^{*+} D_s^{*-}$ channels, but a resonance rather than a bound state is found. The molecular picture is also assumed in Ref.~\cite{Xie:2022lyw} where production rates are evaluated. The molecular picture is not the only suggested possibility. In fact, in Ref.~\cite{Agaev:2022pis} and using also QCD sum rules, a scalar diquark-antidiquark state is studied to explain the $D_s^+ D_s^-$ peak. On the other hand, a state of tetraquark nature is also proposed in Ref.~\cite{Guo:2022crh} using a chromomagnetic interaction model. In Ref.~\cite{Guo:2022zbc}, the state is supposed to be the ordinary $\chi_{c0}(2P)$ state of the constituent quark model. In Ref.~\cite{Badalian:2023qyi}, the $X(3915)$ and $X(3960)$ are considered as four-quark states.

With different views on the subject, it is important to come with new ideas that can help us gain further insight into the origin and meaning of the observed peak and the nature of whatever state is responsible for it. Knowing that: \begin{inparaenum}[\itshape(i)] \item the $X(3915)$ was observed by the Babar collaboration in the $\gamma \gamma \to J/\psi \omega$ reaction \cite{Belle:2009and} with quantum numbers\footnote{It has been argued~\cite{Zhou:2015uva} that the helicity-2 dominance hypothesis, which is reasonable for the coupling of a $2^{++}$ $c\bar c$ state to two photons~\cite{Li:1990sx} and is supported by experimental measurements~\cite{BESIII:2012uyb}, adopted in the BaBar analysis~\cite{BaBar:2012nxg}, was not reliable since the $X(3915)$ may not be a purely $c\bar c$ state. If such an assumption was removed, the data appear more consistent with the assignment of $2^{++}$ to the $X(3915)$~\cite{Zhou:2015uva}.} $J^{PC}=0^{++}$ \cite{BaBar:2012nxg}; \item a supposedly equivalent state was reported by the Belle collaboration at $3943$~MeV in the $B \to J/\psi \omega K$ reaction \cite{Belle:2004lle}; \item the $X_0(3930)$ was found in the $D^+ D^-$ mass distribution of $B^+ \rightarrow D^+ D^- K^+$ \cite{LHCb:2020pxc,LHCb:2020bls}; and \item the $X_0(3960)$ was found in the $D_s^+ D_s^-$ mass distribution in $B^+ \rightarrow D_s^+ D_s^- K^+$ \cite{LHCb:2022vsv,LHCb:2022dvn}, we propose to look into the $J/\psi \omega$ mass distribution of the $B^+ \to J/\psi \omega K^+$ reaction. \end{inparaenum} Actually, data for the $B^+ \to J/\psi \omega K^+$ decay are already available \cite{Andreassi:2014skr}. Although  promising, it could be improved in future LHCb runs. 
A peak around $3930$~MeV in the $J/\psi \omega$ mass distribution in $B^+ \to J/\psi \omega K^+$ is also seen in a BaBar experiment~\cite{BaBar:2010wfc}, but as in the $D^+ D^-$ mass distribution of the LHCb, the peak could be due to a superposition of a $J=0$ and a $J=2$ states.

According to the pictures  in Refs.~\cite{Bayar:2022dqa,Ji:2022vdj} a peak should be seen around $3930\,\MeV$ and no extra peak at  $3960\,\MeV$, since we assume that the $D^+ D^-$ and $D_s^+ D_s^-$ peaks are due to the same state $X_0(3930)$. We can make predictions on the position and shape of the peak and the strength compared to those seen in the $D^+ D^-$ and $D_s^+ D_s^-$ mass distributions. Alternatively, and due to a destructive interference with a background, one could also have a dip. Cases like the latter one are seen in hadron physics, for instance in the $\pi \pi$ cross section in the region of the $f_0(980)$ \cite{Protopopescu:1973sh} 
(see also the general discussion in \cite{Dong:2020hxe}). 

On the theoretical side, Ref.\,\cite{Dai:2018nmw} deals with the $B^+ \to J/\psi \omega K^+$ reaction from a different perspective to the one in this manuscript. Indeed, that work looks for signals of $\overline{D}{}^* D^*$ bound states in that reaction. In the $3920 - 3940\,\MeV$ region there are predictions of $\overline{D}{}^* D^*$ bound states in Ref.\,\cite{Molina:2009ct}, where a state with $I^G\left(J^{PC}\right)=0^+\left(0^{++}\right)$ is found around $3940\,\MeV$ and another one with $0^+\left(2^{++}\right)$ around $3920\,\MeV$, both of them coupling to $J/\psi \omega$ in $S$-wave.\footnote{A $\overline{D}{}^\ast D^\ast$ bound state with $0^+(0^{++})$ quantum numbers well below the corresponding threshold is possible according also to Ref.\,\cite{Ji:2022uie}. A $0^+(2^{++})$ $\overline{D}{}^* D^*$ state is also predicted and studied as a HQSS partner of $X(3872)$ in Refs.~\cite{Nieves:2012tt,Hidalgo-Duque:2012rqv,Albaladejo:2015dsa}, but only very close to the $\overline{D}{}^\ast D^\ast$ thresold.} The results of Ref.\,\cite{Dai:2018nmw} indicated that a broad peak around $3920\,\MeV$ in Ref.\,\cite{Andreassi:2014skr} could be attributed mostly to the $J=2$  $\overline{D}{}^* D^*$ state and a cusp like sharp peak of smaller strength appeared at the  $\overline{D}{}^* D^*$ threshold. The present work is also complementary to the one of Ref.~\cite{Dai:2018nmw}, looking for the contribution of the $\overline{D}D$ and $\overline{D}_s D_s$ states. Given the small mixing of pseudoscalar-pseudoscalar and vector-vector components \cite{Dias:2021upl}, the contributions that we find here are additional to those found in Ref.~\cite{Dai:2018nmw}.

\section{Formalism}

\begin{figure}[tbp]   
  \centering
  \includegraphics[width=6.5cm]{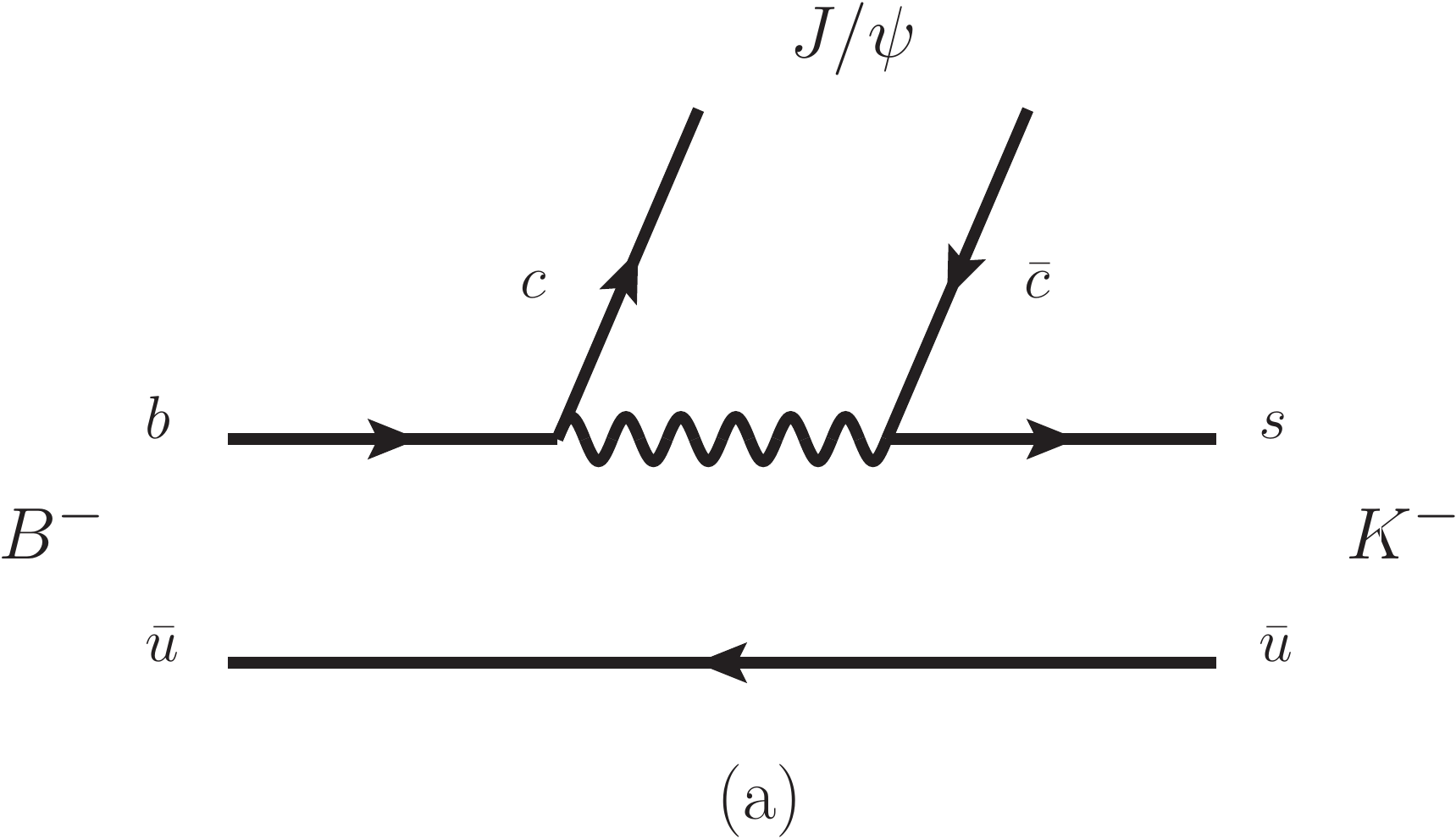} \hspace{8mm}
  \includegraphics[width=6.5cm]{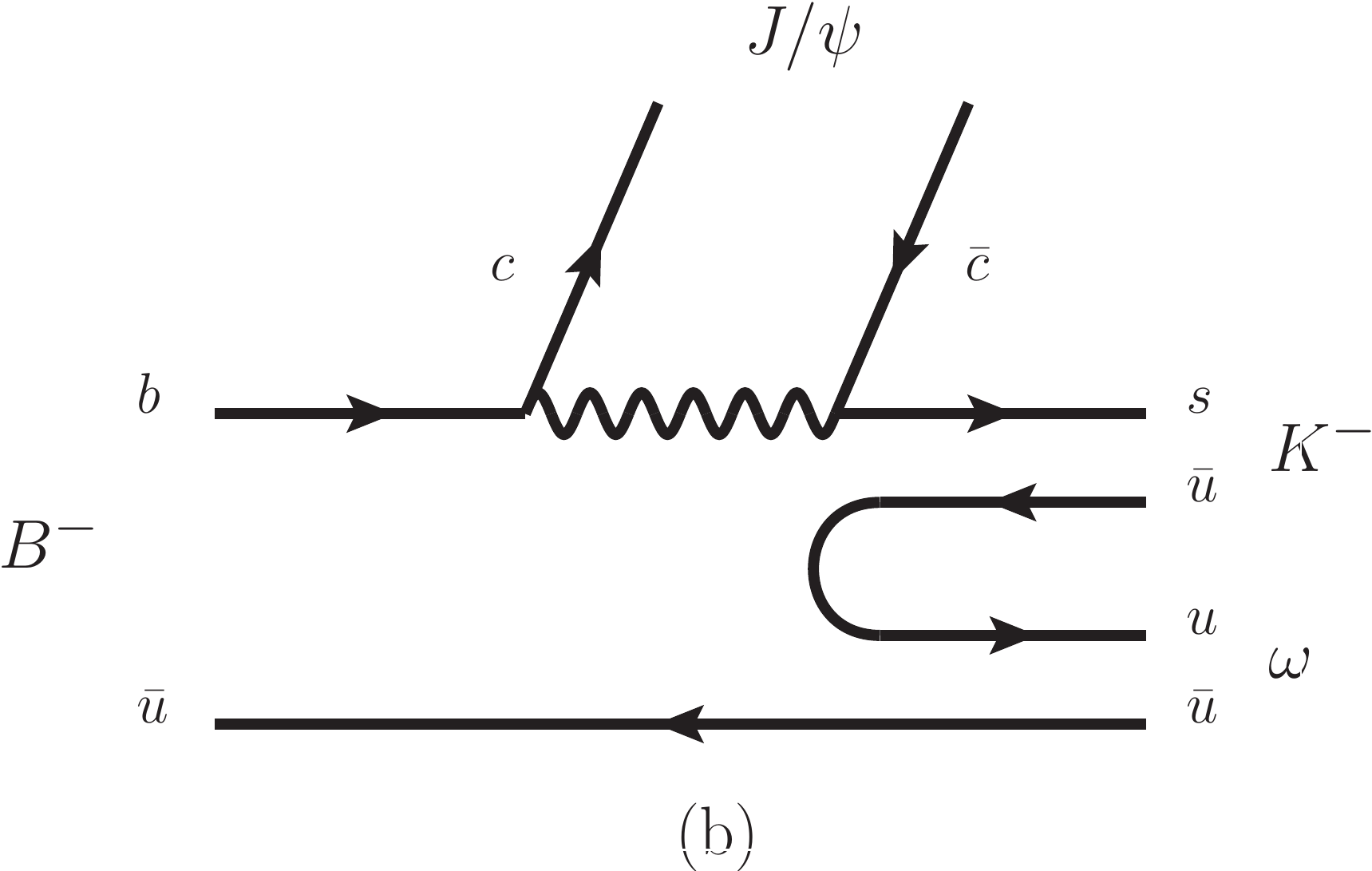}
  \caption{Mechanism at the quark level for the reaction $B^- \to J/\psi \,\omega \, K^- $. (a) Primary quark production with internal emission. (b) Hadronization of the $ s \bar{u} $ component into two mesons.}
  \label{fig1}
\end{figure}

In order to work with $ b $ quarks rather than $ \bar{b} $ we study the charge conjugate reaction, $B^- \to J/\psi \,\omega \, K^- $. We can see in Fig.~\ref{fig1} the mechanism at the quark level for the direct production of the final state without the need of rescattering (tree level). The hadronization is achieved including a $\bar{q} q$ pair with the quantum numbers of the vacuum. Concerning flavor space, we denote $ \bar{q} q  \equiv \sum_{i} \bar{q}_i q_i$ ($i=\left\{u,d,s\right\}$). Then, since we want to get a pseudoscalar meson and a vector, we obtain 
\begin{eqnarray}
 s \bar{u} & \to & \sum_{i} s \bar{q}_i q_i  \bar{u} = (PV)_{31}, 
\label{eq1}
\end{eqnarray}
where $ P, V $ are the  $ q \bar{q} $ matrices in $ \text{SU}(4) $ flavor space written in terms of pseudoscalar or vector mesons:
\begin{equation}
P = \left(
           \begin{array}{cccc}
             \frac{1}{\sqrt{2}}\pi^0 + \frac{1}{\sqrt{3}}\eta + \frac{1}{\sqrt{6}}\eta' & \pi^+ & K^+ & \overline{D}{}^0 \\
             \pi^- & -\frac{1}{\sqrt{2}}\pi^0 + \frac{1}{\sqrt{3}}\eta + \frac{1}{\sqrt{6}}\eta' & K^0 & D^- \\
            K^- & \bar{K}^0 & -\frac{1}{\sqrt{3}}\eta + \sqrt{\frac{2}{3}}\eta' & D_s^-\\
D^0 & D^+ & D_s^+ & \eta_c \\
          \end{array}
         \right),
  \label{eq2}
\end{equation}
\begin{equation}
V = \left(
           \begin{array}{cccc}
             \frac{1}{\sqrt{2}}\rho^0 + \frac{1}{\sqrt{2}}\omega  & \rho^+ & K^{*+} & \overline{D}{}^{\ast 0}  \\
             \rho^- & -\frac{1}{\sqrt{2}}\rho^0 + \frac{1}{\sqrt{2}}\omega  & K^{*0} & D^{\ast -}  \\
            K^{*-} & \bar{K}^{*0} & \phi & D_s^{\ast -} \\
D^{\ast 0} & D^{\ast +} & D_s^{\ast +} & J/\psi \\
           \end{array}
         \right).
   \label{eq3}
\end{equation}
In the matrix $ P $ we have considered the standard $ \eta - \eta^{\prime} $ mixing of Ref.~\cite{Bramon:1994cb}. We could as well have the $ (VP)_{31} $ combination, but since we want a final state with a $ K^- $, the $ PV $ product is the appropriate one. Thus, 
\begin{eqnarray}
 s \bar{u} \to (PV)_{31} = K^- \left( \frac{1}{\sqrt{2}} \rho^0 + \frac{1}{\sqrt{2}} \omega \right) + \ldots
\label{eq4}
\end{eqnarray}
We obtain then the  $ J/\psi \,\omega \, K^- $ component for the tree level background in the $ B^- $ decay. 

\begin{figure}[tbp]   
  \centering
  \includegraphics[width=8cm]{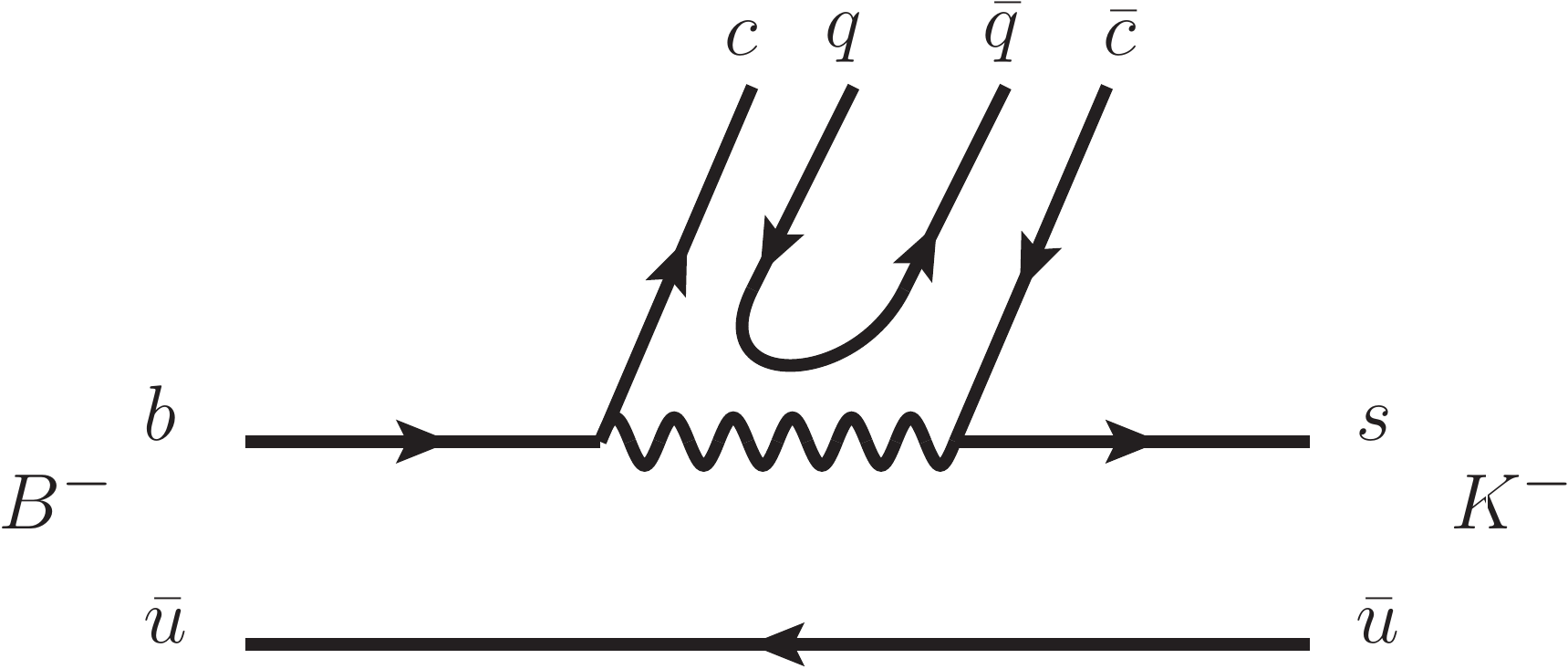}
  \caption{Hadronization of the $ c \bar{c} $ quark pair. }
  \label{fig2}
\end{figure}

\begin{figure}[tbp]   
  \centering
  \includegraphics[width=5.5cm]{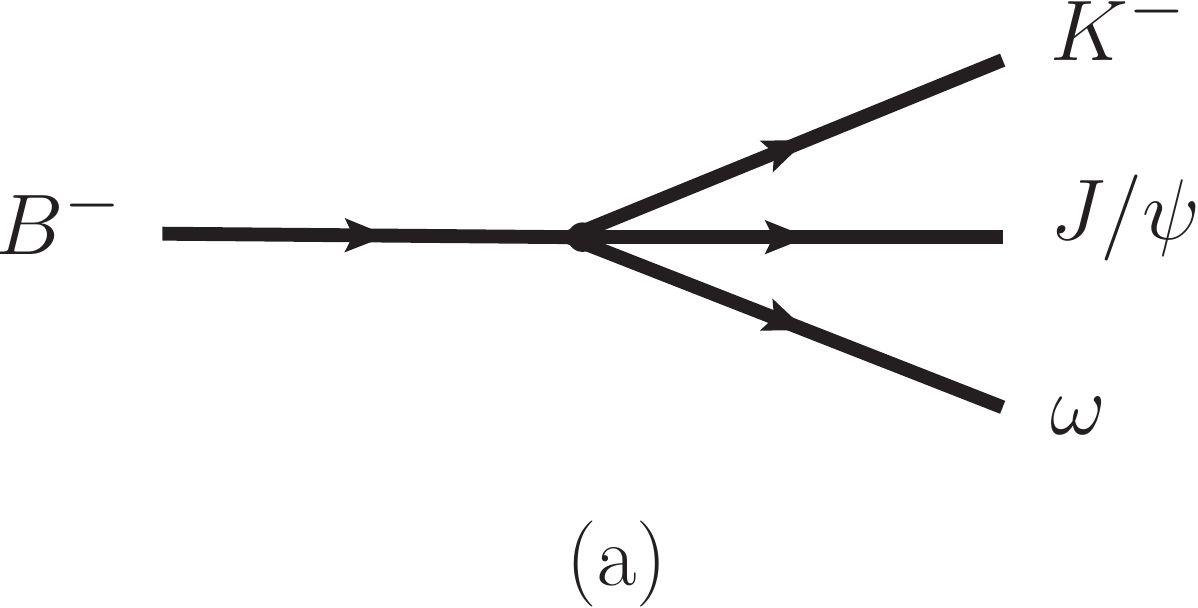} \hspace{8mm}
  \includegraphics[width=8cm]{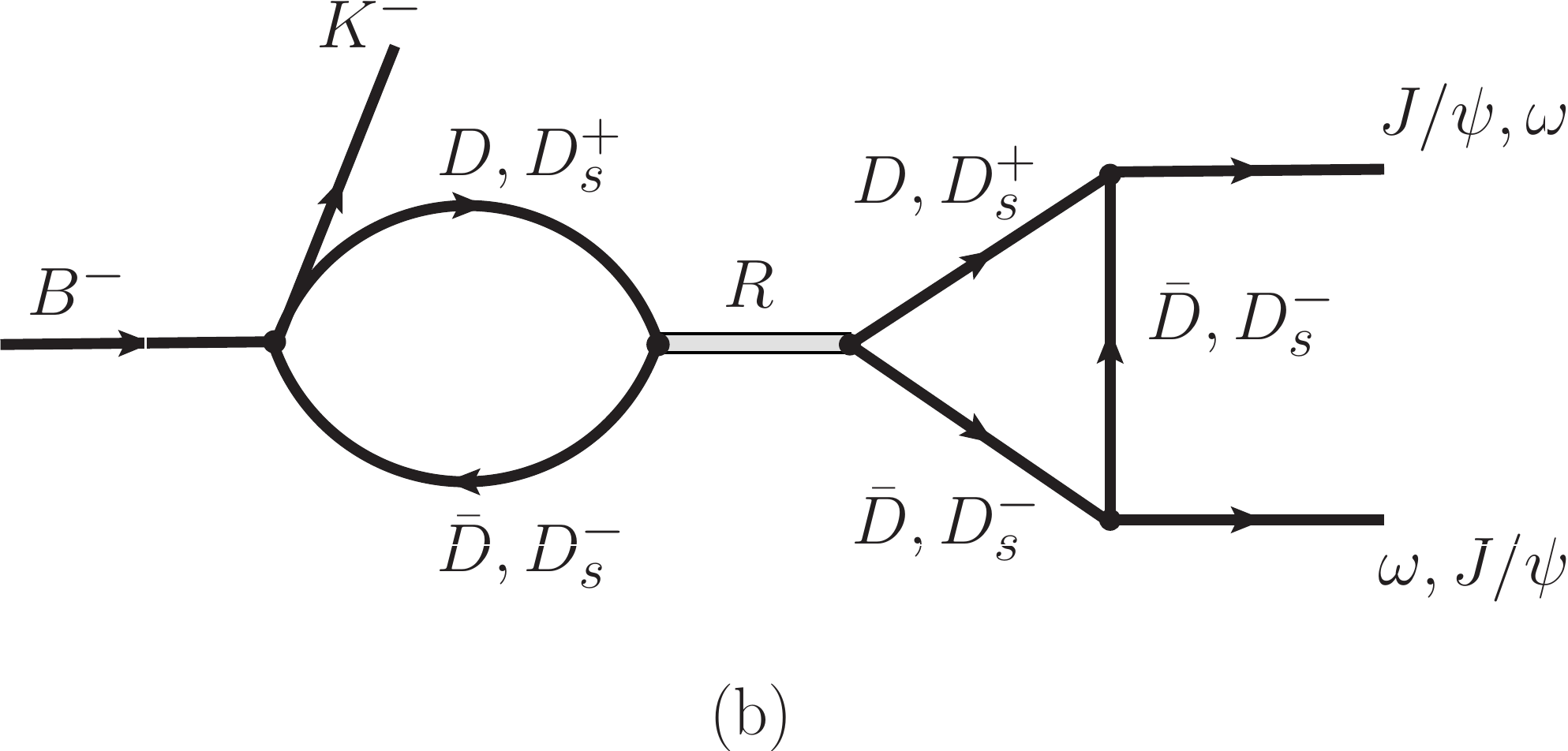} 
  \caption{Mechanisms for the $B^- \to J/\psi \,\omega\,K^-$ decay. (a) Tree-level contribution. (b) Primary $D\bar{D}$, $D_s ^+ D_s ^-$ production followed by rescattering and $ J/\psi \,\omega $ conversion through the triangle diagram. }
  \label{fig3}
\end{figure}

Next, we look for the resonant contribution. The $ D \overline{D} , D_s \overline{D}_s $  components are obtained by hadronizing the $ c \bar{c} $ pair in Fig.~\ref{fig1}(a) as shown in Fig.~\ref{fig2}. We have then: 
\begin{equation}
 c \bar{c} \to \sum_{i} c \bar{q}_i q_i  \bar{c} \to  (P^2)_{44} \nonumber = D^0 \overline{D}{}^0 + D^+ D^- + D_s ^+ D_s ^- + \eta_c \eta_c~.  
\label{eq5}
\end{equation}
We remark that the $ \eta_c \eta_c  $ component does not play any role due to its large mass. The non-resonant (tree-level) and resonant parts of the reaction $B^- \to J/\psi \,\omega \, K^- $ are produced according to the diagrams shown in Fig.~\ref{fig3}. Our isospin phase convention of the $ D$ and $\overline{D}$ doublets is the following: $ D \equiv (D^+, - D^0)$ and $\overline{D}  \equiv  ( \overline{D}{}^0, D^-)$. Then, it follows that the $ I=0 $ state is given by 
\begin{eqnarray}
|D \overline{D}, I=0 \rangle = \frac{1}{\sqrt{2}}(D^{+}D^{-}+D^{0}\overline{D}{}^{0}).             
\label{eq6}
\end{eqnarray}
We associate the weights $ C $ and $ C^{\prime} $ to the mechanisms depicted in Figs.~\ref{fig1}(b) and 2, respectively. Since both of them come from internal emission and hadronization, we will assume that $ C $ and $ C^{\prime} $ are of the same order of magnitude, but the relative sign is not fixed. In addition, we consider the process to proceed via $S$-wave, which is allowed for  $B^- \to J/\psi \,\omega \, K^- $ and  $B^- \to  D \, \overline{D}\,K^- \, / \,  D_s ^+ \, D_s ^- \, K^- $ decays. Therefore, the production amplitude is given by
\begin{align}
\tilde{t}_{J/\psi \,\omega} & =  \frac{1}{\sqrt{2}} C \, \vec{\epsilon}_{J/\psi}\cdot \vec{\epsilon}_{\omega}   \nonumber \\
& + C^{\prime} \Bigg( 
\left[ \sqrt{2} G_{ D \overline{D}}(M_{\rm inv})\, T_{ D \overline{D}, D \overline{D}} (M_{\rm inv}) +  G_{ D_s ^+ D_s ^- }(M_{\rm inv})\, T_{ D_s ^+ D_s ^-, D \overline{D}} (M_{\rm inv}) \right] 
 \left( \frac{1}{\sqrt{2}} \widetilde{V}_{D^{+}D^{-}} + \frac{1}{\sqrt{2}} \widetilde{V}_{D^{0} \overline{D}{}^{0}} \right) \label{eq7} \\
 & 
 \hphantom{C^{\prime} \Bigg( } 
 + 
 \left[ \sqrt{2} G_{ D \overline{D}}(M_{\rm inv})\, T_{ D \overline{D}, D_s ^+ D_s ^- } (M_{\rm inv}) +  G_{ D_s ^+ D_s ^- }(M_{\rm inv})\, T_{ D_s ^+ D_s ^-, D_s ^+ D_s ^- } (M_{\rm inv}) \right]  \widetilde{V}_{ D_s ^+ D_s ^-} \Bigg)~,
   \nonumber
\end{align}
where the $ \vec{\epsilon}_i $ ($i=\left\{ J/\psi, \omega \right\} $) are the polarization vector of the vector mesons, $ G_{ D \overline{D}}(M_{\rm inv})$ and $ G_{ D_s ^+ D_s ^- }(M_{\rm inv}) $ are respectively the loop functions of the $  D \overline{D}$ and $D_s ^+ D_s ^-  $ intermediate states, depicted in Fig.~\ref{fig3}(b), and $ M_{\rm inv} $ is the invariant mass of the $  D \overline{D} $ or  $ D_s ^+ D_s ^-  $ system. In addition, $ T_{i,j} $ represent the elements of the  unitarized transition matrix between $  D \overline{D}, D_s ^+ D_s ^-  $  states, obtained in Ref.~\cite{Bayar:2022dqa} from 
\begin{eqnarray}
      T=\left[\mathbb{I}-VG\right]^{-1}V~,   \label{eq8}
\end{eqnarray} 
where $ V $ here represents the interaction potential matrix. We use the same $ G $ and $ T $ matrices as in Ref.~\cite{Bayar:2022dqa}. It is worth mentioning that even if throughout the manuscript we are apparently using a $\text{SU}(4)$ formalism, one is actually only making use of $\text{SU}(3)$ symmetry and the $ q \bar{q} $ character of the meson states~\cite{Sakai:2017avl}.

On the other hand, the functions $  \widetilde{V}_{j} \, ( j = D^{+}D^{-}, D^{0} \overline{D}{}^{0}, D_s ^+ D_s ^- )$ correspond to the triangle loops in Fig.~\ref{fig3}(b), with  $D^{+}D^{-}$, $D^{0} \overline{D}{}^{0}$, and $ D_s ^+ D_s ^-  $ being the intermediate states. We shall see that the $\widetilde{V}_j $ also contains the $\vec{\epsilon}_{J/\psi}\cdot \vec{\epsilon}_{\omega} $ factor.  Before the evaluation of their magnitudes, it is easy to see that the $ D_s ^+ D_s ^-  $ channel does not contribute. Indeed, it would necessarily involve a $ D_s ^+ D_s ^- \omega $ vertex which is zero in the scheme of Ref.~\cite{Bayar:2022dqa}, since the $ \omega $ meson has only light-flavour quarks and has no overlap with the charm and strange content of the $ D^+_s D^-_s$ pair. As a consequence, $\tilde{t}_{J/\psi \,\omega} $ in Eq.~(\ref{eq7}) gets simplified and can be rewritten as 
\begin{align}
\tilde{t}_{J/\psi \,\omega} & =  \frac{1}{\sqrt{2}} C \, \vec{\epsilon}_{J/\psi}\cdot \vec{\epsilon}_{\omega}   \nonumber \\
& + C^{\prime} \Bigg( 
\left[ \sqrt{2} G_{ D \overline{D}}(M_{\rm inv})\, T_{ D \overline{D}, D \overline{D} } (M_{\rm inv}) +  G_{ D_s ^+ D_s ^- }(M_{\rm inv})\, T_{ D_s ^+ D_s ^-, D \overline{D}} (M_{\rm inv}) \right] 
 \left( \frac{1}{\sqrt{2}} \widetilde{V}_{D^{+}D^{-}} + \frac{1}{\sqrt{2}} \widetilde{V}_{D^{0} \overline{D}{}^{0}} \right) \Bigg)~.\label{eq9} 
\end{align}

\begin{figure}[tbp]   
  \centering
 \includegraphics[width=5.5cm]{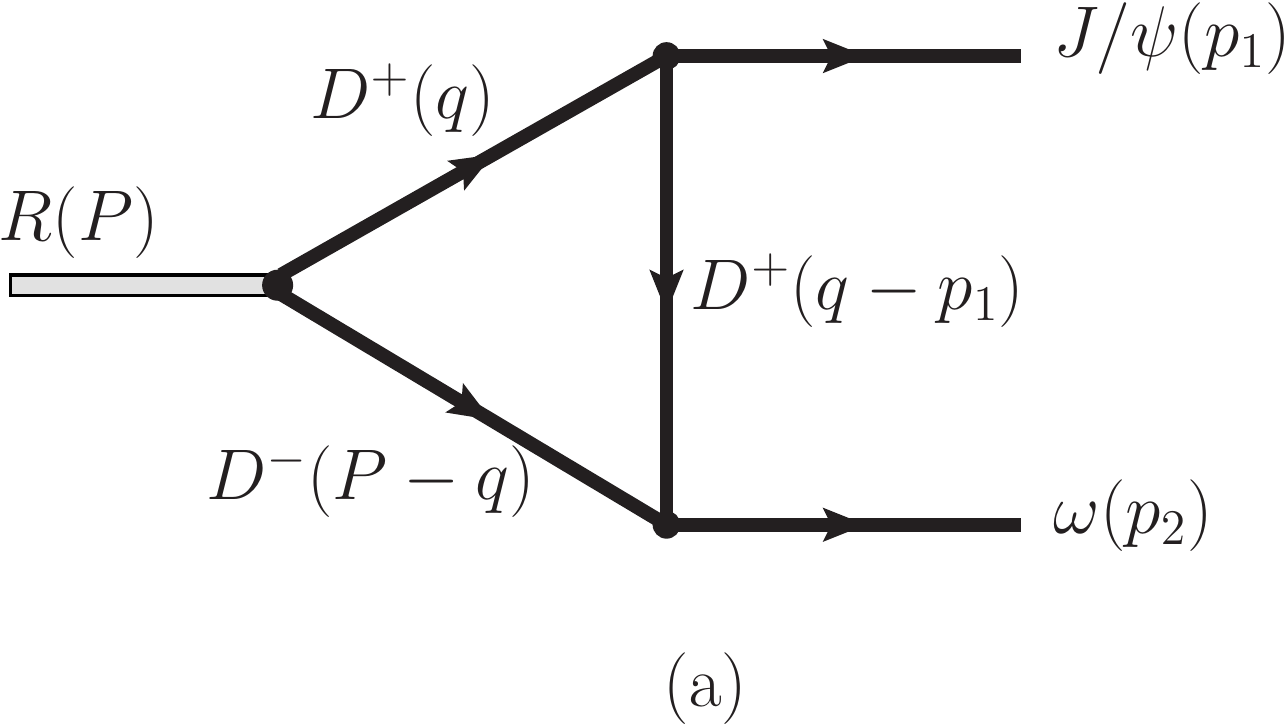} \hspace{8mm}
 \includegraphics[width=5.5cm]{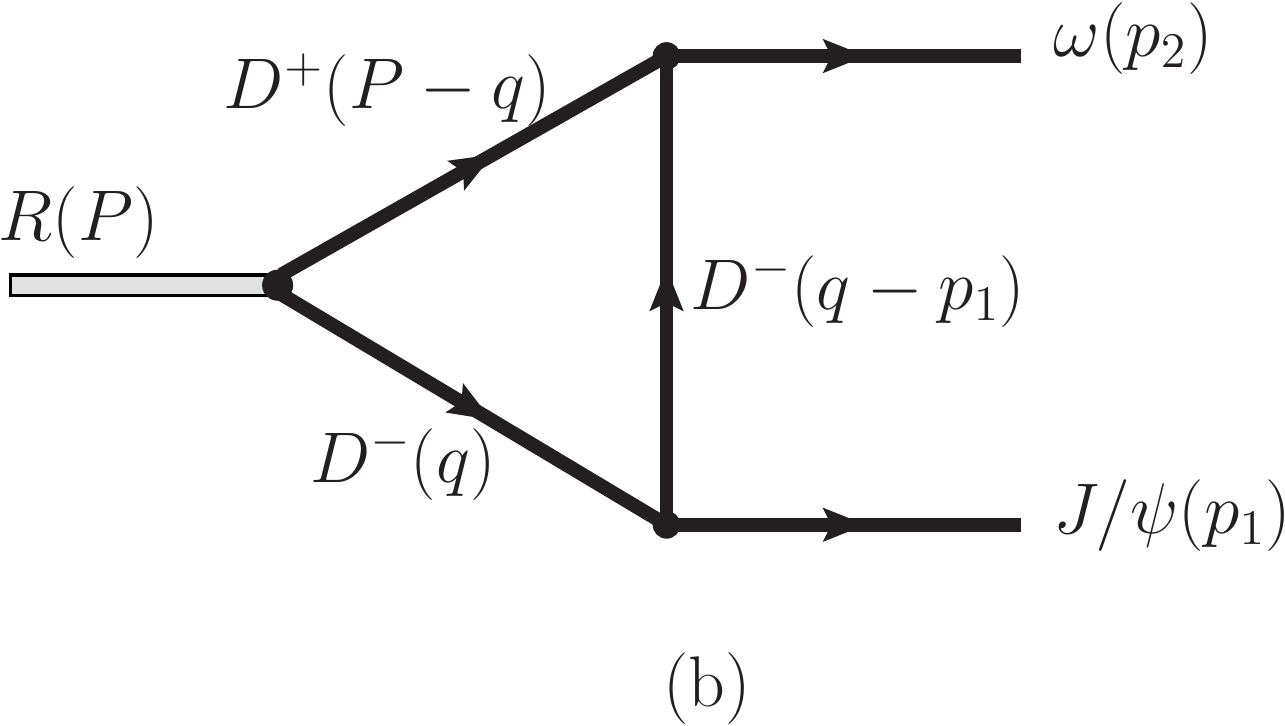}
  \caption{Triangle diagram with $  D^{+}D^{-} $ intermediate state. The panels (a) and (a) are essentially the same but with $ J/\psi $ ad  $\omega $ exchanged. }
  \label{fig4}
\end{figure}

In order to calculate the triangle loop function, we show in Fig.~\ref{fig4} the diagram with the explicit momenta. The analytical expression for diagram (a) reads:
\begin{eqnarray}
 -i \widetilde{V}_{D^{+}D^{-}}^{(a)} & = & - i  \left( \frac{m_{D^*}}{m_{K^*}}\right)^2 \int \frac{d^4q}{(2\pi)^4}
 \frac{i^3 \left( -i V_1 \right) \left(-i V_2 \right) \, \thetaCO (q_\text{max} - |\vec{q}|) }{\left[ q^2-m^2_{D}+i\varepsilon \right] \left[ (P-q)^2-m^2_{D}+i\varepsilon \right] \left[ (q-p_1)^2-m^2_{D} \right]}~,
    \label{eq10}
\end{eqnarray}
where $ V_1, V_2 $ correspond to $  J/\psi D^{+}D^{-} $ or $  \omega D^{+}D^{-} $ vertices, which are evaluated through the following Lagrangian:
\begin{eqnarray}
    \mathcal{L}_{\mathrm{VPP}} & = & -i g\left\langle\left[P, \partial_{\mu} P\right] V^{\mu}\right\rangle,    
     \label{eq10a}
\end{eqnarray}
with $P,V$ denoting the matrices of Eqs.~(\ref{eq2})-(\ref{eq3}) respectively, $ \left\langle\ \cdots \right\rangle $ the trace over the flavor space, and $ g $ the coupling $g=m_V/(2f_\pi)$ ($m_V=800$ MeV, $f_\pi=93$ MeV). The factor $ \thetaCO (q_\text{max} - |\vec{q}|) $ appears due to the sharp cutoff regularization used in the $ T_{ij} $ matrix, which takes the form   $T_{ij} (q, q^{\prime} ) = T_{ij} \, \thetaCO (q_\text{max} - |\vec{q}|) \, \thetaCO (q_\text{max} - |\vec{q^{\prime}}|)$ \cite{Gamermann:2009uq,Song:2022yvz}. Here, we take $ q_\text{max} = 750\,\MeV$ as in Ref.~\cite{Bayar:2022dqa}. The factor $  (m_{D^*}/m_{K^*})^2 $ in Eq.~\eqref{eq10} is a normalization stemming from the different weight factors $ 1/\sqrt{2\omega}$ of the light and heavy mesons~\cite{Liang:2014eba}. On the other hand, it can also be seen that the contributions of diagrams in Fig.~\ref{fig4} [(a) and (b)] are the same, and that $ \widetilde{V}_{D^{+}D^{-}} = \widetilde{V}_{D^{0} \overline{D}{}^{0}} $. 
Hence, we have
\begin{eqnarray}
 \widetilde{V} & = &  \widetilde{V}_{D^{+}D^{-}}^{(a+b)} + \widetilde{V}_{D^{0} \overline{D}{}^{0}}^{(a+b)} 
 =4  \widetilde{V}_{D^{+}D^{-}}^{(a)}  ,
    \label{eq11}
\end{eqnarray}
with 
\begin{eqnarray}
\widetilde{V}_{D^{+}D^{-}}^{(a)} & = & i  \left( \frac{m_{D^*}}{m_{K^*}}\right)^2 \int \frac{d^4q}{(2\pi)^4}
 \frac{\displaystyle \left( -2 g \, \epsilon_{J/\psi} \cdot q \right) \left( \frac{g}{\sqrt{2}} \, \epsilon_{\omega} \cdot (2q-P-p_1) \right) \thetaCO \left( q_\text{max} - \left\lvert \vec{q} \right\rvert \right) }{\left[ q^2-m^2_{D}+i\varepsilon \right] \left[ (P-q)^2-m^2_{D}+i\varepsilon \right] \left[ (q-p_1)^2-m^2_{D} \right]}~,
    \label{eq12}
\end{eqnarray}
One can take advantage of the fact that the $ D $-meson exchanged between the $ J/\psi $ and $ \omega $ in Fig.~\ref{fig4} is far off-shell and we can factor it out of the integral. Then, we perform the $ q^0 $ integration analytically in Eq.~(\ref{eq12}) via Cauchy's theorem, taking the pole of the $ q $ line, and thus $ q $ is placed on-shell. Therefore, 
\begin{equation}
 \frac{1}{(q-p_1)^2-m^2_{D} } = \frac{1}{q^2 + p_1^2 - 2 q \cdot p_1 - m^2_{D} } \simeq \frac{1}{m^2_{J/\psi} - 2 m_{J/\psi} m_{D} },
    \label{eq13}
\end{equation}
where we have considered that $ J/\psi $ and the intermediate $D$ on-shell come with relative small momentum. As a result, we obtain
\begin{eqnarray}
 \widetilde{V} & \simeq & - \frac{8 g^2}{\sqrt{2}}  \left( \frac{m_{D^*}}{m_{K^*}}\right)^2 \frac{1}{m^2_{J/\psi} - 2 m_{J/\psi} m_{D} }  \times  \int \frac{d^3q}{(2\pi)^3}
\frac{1}{\omega_D(q)} \frac{\vec{\epsilon}_{J/\psi} \cdot \vec{q} \, \vec{\epsilon}_{\omega}\cdot  (2 \vec{q}-\vec{p_1}) \thetaCO (q_\text{max} - |\vec{q}|)}{(P^0)^2 - 4 \omega_D^2(q) + i\varepsilon } 
 \nonumber \\
 & = & - \frac{16 g^2}{3\sqrt{2}}  \left( \frac{m_{D^*}}{m_{K^*}}\right)^2 \frac{1}{m^2_{J/\psi} - 2 m_{J/\psi} m_{D} } 
\vec{\epsilon}_{J/\psi} \cdot \vec{\epsilon}_{\omega} 
 \times \int_{|\vec{q}| < q_\text{max}} \frac{d^3q}{(2\pi)^3} \frac{1}{\omega_D(q)} \frac{\vec{q}\,^2}{(P^0)^2 - 4 \omega_D^2(q) + i\varepsilon } ,
    \label{eq1415}
\end{eqnarray}
where $ P^0 = M_{\rm inv} $ and $ \omega_{D} (q) \equiv \sqrt{\vec{q}\,^2 + m_{D} ^2 }$. In \ref{app:integral} we discuss the integral appearing in Eq.~\eqref{eq1415}. So, by using 
\begin{eqnarray}
 \widetilde{V} & \equiv &  \widetilde{V}^{\prime} \, \vec{\epsilon}_{J/\psi} \cdot \vec{\epsilon}_{\omega} , 
  \nonumber \\ 
\tilde{t}_{J/\psi \,\omega}   & \equiv & \tilde{t}_{J/\psi \,\omega}^{\prime}  \, \vec{\epsilon}_{J/\psi} \cdot \vec{\epsilon}_{\omega}   ,
    \label{eq16}
\end{eqnarray}
we can now write
\begin{equation}
\tilde{t}_{J/\psi \,\omega}^{\prime} =   \frac{1}{\sqrt{2}} C + C^{\prime} \left[ \sqrt{2} G_{ D \overline{D}}(M_{\rm inv}) T_{ D \overline{D}, D \overline{D}} (M_{\rm inv}) +  G_{ D_s ^+ D_s ^- }(M_{\rm inv}) T_{ D_s ^+ D_s ^-, D \overline{D}} (M_{\rm inv}) \right] \frac{\widetilde{V}^{\prime}}{\sqrt{2}}~. 
             \label{eq17}
\end{equation}
The mass distribution is obtained from the expression:
\begin{eqnarray}
  \frac{d\Gamma}{dM_{\rm inv}}=\frac{3}{(2\pi)^3}\frac{1}{4 m^2_{B}} p_{K^-} \tilde {p}_{\omega} | \tilde{t}_{J/\psi \,\omega}^{\prime} |^2,
      \label{eq18}
\end{eqnarray}
where the factor 3 comes from summing over the $ J/\psi$ and $\omega$ polarizations, and  
\begin{eqnarray}
    p_{K^-}&= & \frac{\lambda^{1/2}(m_{B}^2,m^2_{K},M_{\rm inv}^2)}{2m_{B}},  \label{eq19}    \\
    \tilde{p}_{\omega} & = & \frac{\lambda^{1/2}(M_{\rm inv}^2,m_{J/\psi}^2,m_{\omega}^2)}{2M_{\rm inv}}. 
    \label{eq20}
\end{eqnarray}

\section{Partial decay width of the $ X_0(3930) $ to $ J/\psi \, \omega $}

In Ref.~\cite{Bayar:2022dqa} the $ X_0(3930) $ state came from the coupled channels $  D \overline{D} $ and $D_s ^+ D_s ^-  $, mostly  $D_s ^+ D_s ^-  $. The width originating from its decay to the open $  D \overline{D} $ channel was found  to be $13\,\MeV$. Here we benefit from this previous evaluation to obtain the decay width of  $ X_0(3930) $ to $ J/\psi \, \omega $. 

We can take the diagrams of Fig~\ref{fig4} and together with the $ D^{0} \overline{D}{}^{0} $ intermediate diagram we find an effective vertex of the resonance and $ J/\psi \, \omega $ as 
\begin{eqnarray}
    g_{R, J/\psi \, \omega}&= & g_{R, D \overline{D}} \frac{\widetilde{V}^{\prime}}{\sqrt{2}} \vec{\epsilon}_{J/\psi} \cdot \vec{\epsilon}_{\omega} 
 \equiv  g_{R,J/\psi \, \omega}^{\prime} \vec{\epsilon}_{J/\psi} \cdot \vec{\epsilon}_{\omega}  ,  \label{eq21}
\end{eqnarray}
where $ \widetilde{V}^{\prime} $ has been defined in Eqs.~(\ref{eq1415}) and~(\ref{eq16}), and $ g_{R, D \overline{D}} $ is the coupling of the resonance to $  D \overline{D} $ found in Ref.~\cite{Bayar:2022dqa}, 
\begin{equation}
 g_{R, D \overline{D}} = -1.00 - i 2.71\,\eV[G]~.
  \label{eq22}
\end{equation}
The  $ X_0(3930) $ width is then given by 
\begin{eqnarray}
 \Gamma_{R}^{J/\psi \, \omega} = \frac{3}{8\pi}\frac{1}{ m^2_{R}} \tilde {p}_{\omega} | g_{R, J/\psi \,\omega}^{\prime} |^2,
  \label{eq23}
\end{eqnarray}
where the factor 3 comes again from the sum of $  J/\psi , \omega $ polarizations. 
For the sake of coherence, we use the mass $ m_{R} = 3932.72 $ MeV from Ref.~\cite{Bayar:2022dqa}.

\section{Further considerations}

\begin{figure}[tbp]   
  \centering
  \includegraphics[width=6cm]{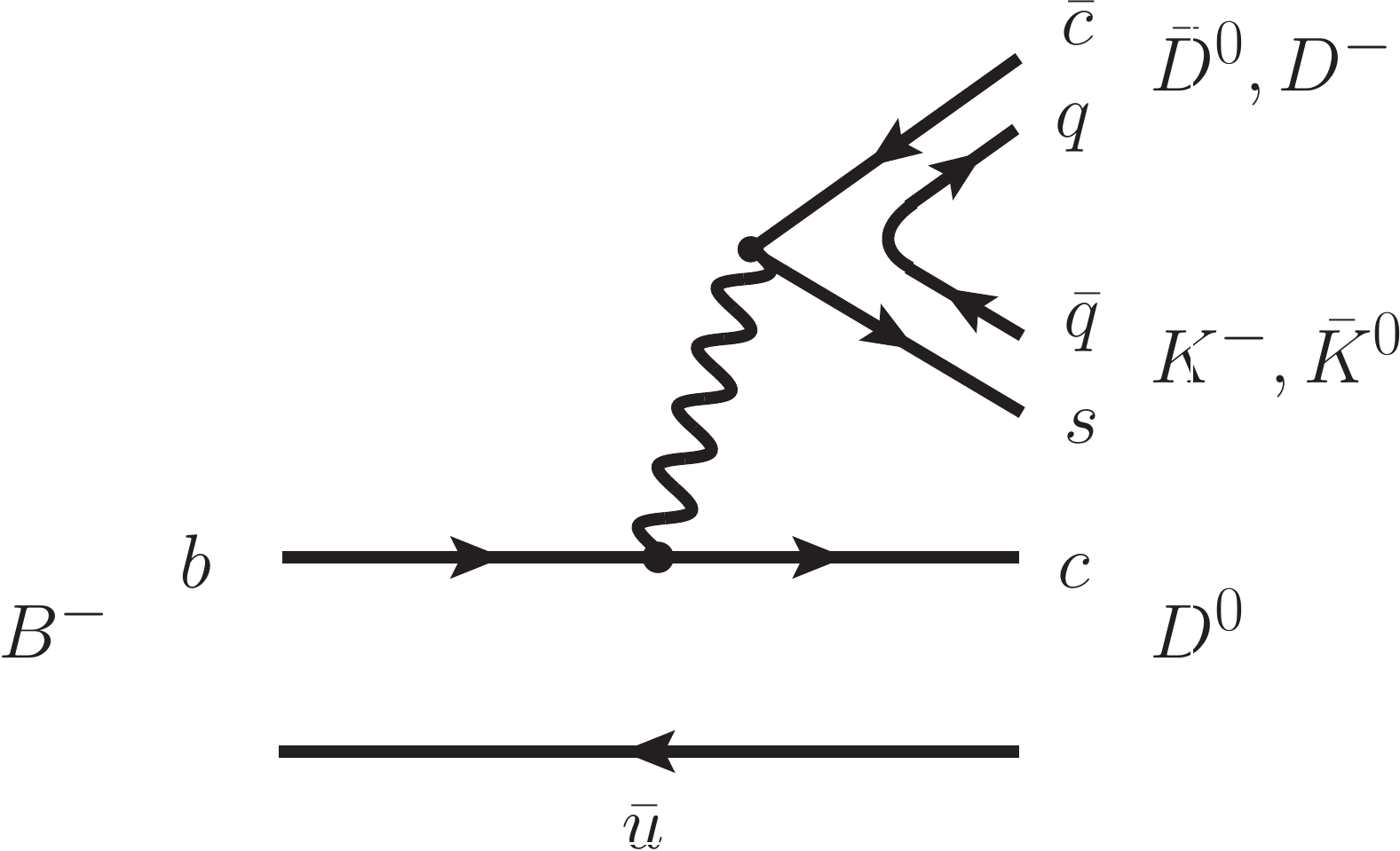}
  \caption{Topology for $B^- \to J/\psi \,\omega \, K^- $ production based on external emission and hadronization. }
  \label{fig5}
\end{figure}

As mentioned above, in Ref.~\cite{Dai:2018nmw} the same reaction $B^- \to J/\psi \,\omega \, K^- $ was studied but looking for states of  $  D^* \overline{D}{}^* $ nature. The same quark topologies as in Figs.~\ref{fig1} and~\ref{fig2} were considered but, in addition, a diagram corresponding to external emission, shown here in Fig.~\ref{fig5}, was also taken into account. The hadronization of the $ s \bar{c} $ pair gives rise to: 
\begin{equation}
s  \bar{c} \to \sum_{i} s \bar{q}_i q_i \bar{c} =  (P^2)_{34}  = K^- \overline{D}{}^0 +  \overline{K}{}^0 D^-  - \frac{1}{\sqrt{3}} \eta D_s^- +\sqrt{ \frac{2}{3} } \eta^\prime D_s^- + \eta_c D_s^-~.\label{eq24}
\end{equation}
Among all these terms, there is one that can lead to our resonance, namely $  K^- \overline{D}{}^0  $. Together with the $ D^0  $ meson of the  $ c \bar{u} $ pair in Fig.~\ref{fig5} we obtain the term 
\begin{eqnarray}
H \equiv K^- D^0 \overline{D}{}^0  . 
\label{eq25}
\end{eqnarray}
Considering Eqs.~(\ref{eq6}) and~(\ref{eq7}) thus will lead to an extra term,  
\begin{eqnarray}
 C^{\prime \prime} \frac{1}{\sqrt{2}} G_{ D \overline{D}}(M_{\rm inv}) T_{ D \overline{D}, D \overline{D}} (M_{\rm inv}) \frac{1}{\sqrt{2}} \widetilde{V}
\nonumber 
\end{eqnarray}
Hence, the term mentioned above can be incorporated by performing the following replacement in Eq.~(\ref{eq7}), 
\begin{equation}
C^{\prime} \sqrt{2} G_{ D \overline{D}}(M_{\rm inv}) T_{ D \overline{D}, D \overline{D}} (M_{\rm inv}) \to \left( C^{\prime}  \sqrt{2} + \frac{1}{\sqrt{2}}  C^{\prime \prime} \right) G_{ D \overline{D}}(M_{\rm inv}) T_{ D \overline{D}, D \overline{D}} (M_{\rm inv})~.
\end{equation}
 Since the external emission topology is color favored, we can expect that $  C^{\prime \prime} \simeq  3 C^{\prime } $. As a consequence, the new mechanism can give about the same contribution as the former intermediate $ D \overline{D} $ state contribution. Yet, since the resonance that we consider couples mostly to $ D_s ^+ D_s ^- $ and is close to the $ D_s ^+ D_s ^- $ threshold, both the $ G_{ D_s ^+ D_s ^- } $ and  $ T_{ D_s ^+ D_s ^-, D \overline{D}  } $  have larger strength than $ G_{ D \overline{D}} $ and $ T_{ D \overline{D}, D \overline{D}}  $, and the new $ D \overline{D} $ contribution, as well as the former one, will be small compared to the contribution of the intermediate $ D_s ^+ D_s ^- $ state. In Ref.~\cite{Dai:2018nmw} the largest contribution was obtained from a $ D^\ast \overline{D}{}^\ast$ $(J^{PC} = 2^{++})$ state. On experimental grounds, in order to isolate the $ D \overline{D}$ ,  $D_s ^+ D_s ^- $ state, the $2^{++}$ contribution can be identified either by a partial wave analysis or by using the moments method~\cite{LHCb:2014ioa,Du:2017zvv,Mathieu:2019fts,Bayar:2022wbx}, projecting the mass distribution as 
\begin{eqnarray}
\int   \frac{d^2\Gamma}{dM_{\rm inv}d\Omega} (M_{\rm inv},\cos{\theta})Y_{i0}(\cos{\theta})d\Omega,  
      \label{eq28}
\end{eqnarray}
where $ \theta $ is the angle between the $ K^- $ and $ \omega $ in the $  J/\psi \,\omega $ rest frame, which is the same as the angle between $ \omega $ in the $  J/\psi \,\omega $ rest frame and  $ K^- $ in the $ B^- $ rest frame; $ \Omega $ is the solid angle of  $  J/\psi \,\omega $ in their rest frame.

\section{Results}

In Ref.~\cite{Bayar:2022dqa} the dimensional regularization was used, and accordingly two subtraction constants were introduced: one in each of the $  D \overline{D} $ and $  D_s ^+ D_s ^- $ channels. Notwithstanding this, one can find the equivalent cutoff values by demanding that the loop $ G $ functions are the same at their respective thresholds. In this way, we obtain $ q_\text{max} = 550\,\MeV$ for $  D \overline{D} $ and $ 1070\,\MeV$ for $   D_s ^+ D_s ^- $. Since we are concerned about the state coupling mostly to $  D_s ^+ D_s ^- $, the use of a different $ q_\text{max} $ for the $  D \overline{D} $ channel does not change much the results for that state. We need $ q_\text{max} $ for the triangle loop function and we shall take an intermediate value, \textit{i.e.} $ q_\text{max} = 750\,\MeV$. We shall quantify the uncertainties concerning small changes of that parameter. 

\begin{figure}
\centering
  \includegraphics[width=0.47\textwidth,keepaspectratio]{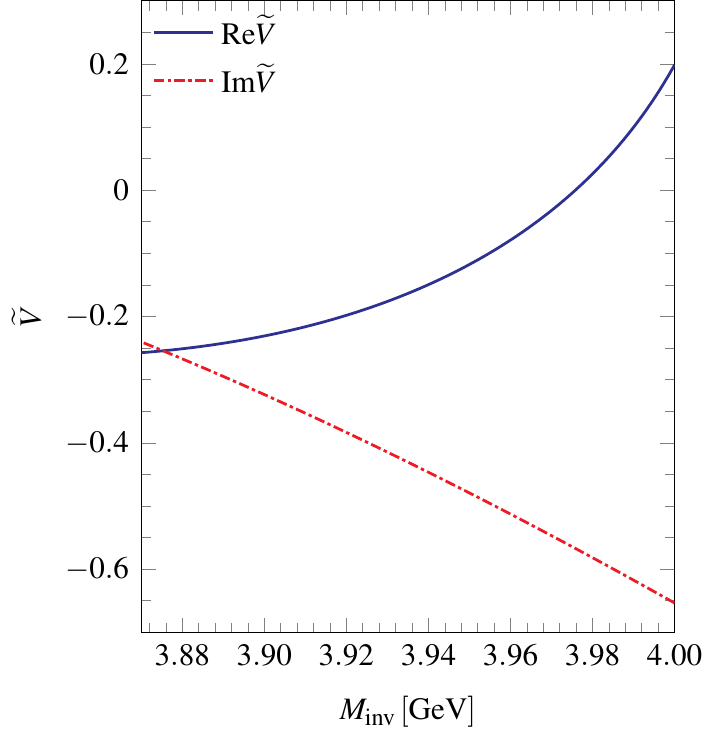}
\caption{Real and imaginary parts of triangle loop contribution $ \widetilde{V}^{\prime} $ given in Eqs.~(\ref{eq1415}) and~(\ref{eq16}). }
\label{fig6}
\end{figure}

Let us start the discussion of the results by evaluating the triangle contribution $ \widetilde{V}^{\prime} $. As we can see in Eqs.~(\ref{eq1415}) and~(\ref{eq16}), $ \widetilde{V}^{\prime} $ looks like the ordinary meson-meson $ G $ function, except for the extra $ \vec{q}\,^2 $ factor in the numerator (see also \ref{app:integral}). Since the threshold of $  D \overline{D} $ is at $ 3730\,\MeV $, in the region of $ 3870\,\MeV $ to $ 4000\,\MeV $ it should behave like a $ G $ function above threshold where the real part increases from a negative number and the imaginary part is negative and grows in magnitude when the energy increases. This is what we observe in Fig.~\ref{fig6}. The magnitude of $ \widetilde{V}^{\prime} $  around $ 3930\,\MeV $ is of the order of $ (2\text{ -- }3) \times 10^{-1} $.

\begin{figure}
\centering
  \includegraphics[width=0.47\textwidth,keepaspectratio]{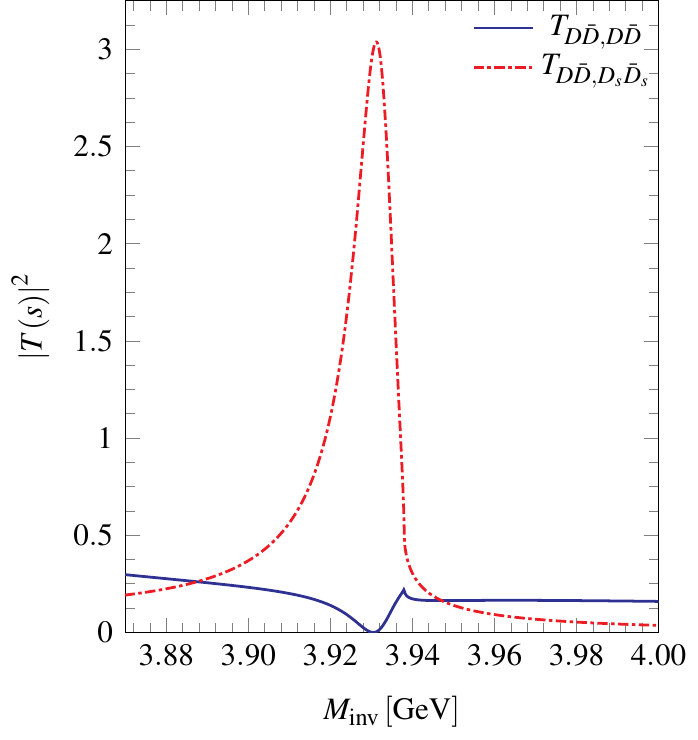}
\caption{Squared  modulus of the transition matrix elements $ T_{ D \bar{D}, D \bar{D}} $ and $ T_{ D_s ^+ D_s ^-, D \bar{D}} $ appearing in $ \tilde{t}_{J/\psi \,\omega} $ in Eq.~(\ref{eq9}). }
\label{fig7}
\end{figure}

For completeness, in Fig.~\ref{fig7} we plot the squared  modulus of  the elements of the unitarized transition matrix $ T $ that are relevant in the estimation of the $ \tilde{t}_{J/\psi \,\omega} $ in Eq.~(\ref{eq9}). As we discussed previously, in the amplitude of Eq.~(\ref{eq9}) we needed the $ T_{ D \overline{D}, D \overline{D}} $ and $ T_{ D_s ^+ D_s ^-, D \overline{D}} $ transitions, and in Fig.~\ref{fig7} it can be clearly observed that $ T_{ D_s ^+ D_s ^-, D \overline{D}} $ is the most relevant transition.

\begin{figure}
\centering
\includegraphics[width=0.47\textwidth,keepaspectratio]{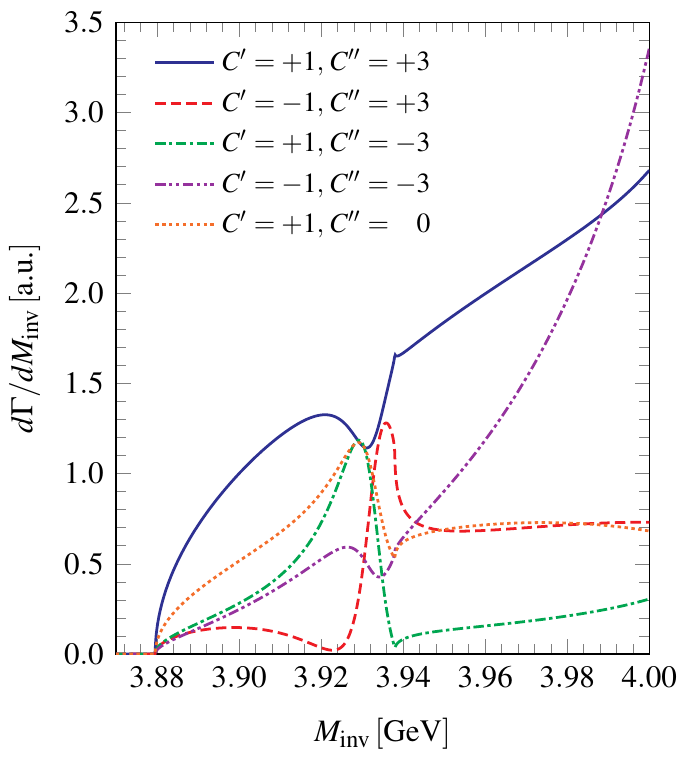}
\caption{Differential decay for the reaction $B^+ \to K^+ J/\psi \, \omega $, as a function of the  $ J/\psi \omega $ mass distribution, denoted as $ \sqrt{s} $, with different values of the weights $C, C^{\prime}, C^{\prime \prime}$ appearing in $\tilde{t} _{J/\psi \omega}$. }
\label{fig8}
\end{figure}

We plot in Fig.~\ref{fig8} the mass distribution $ d \Gamma / d M_\text{inv} $ for  $ J/\psi \omega  $ production in the reaction $B^- \to K^- J/\psi \,\omega $. We consider different sets of weights $\left\{ C, C^{\prime}, C^{\prime \prime} \right\}$, present in $\tilde{t} _{J/\psi \omega}$, based on our estimates of the strength of internal and external emissions. We take $C $ and $ C^{\prime}$ of the order of 1 an $ C^{\prime \prime} $ of the order of 3, but play with possible relative signs of the magnitudes. 
Further experimental works will give us information on which set should be better to reproduce experimental data. However, these estimations suggest interesting findings. In the region of $3920-3950\,\MeV$ an interference appears among the tree-level and resonant terms present in the amplitude  $ \tilde{t}_{J/\psi \,\omega} $, which might be constructive or destructive, depending on the set of parameters utilized. 
One should recall that interferences leading to dips in amplitudes or mass distributions are common when dealing with coupled channels~\cite{Dong:2020hxe}.
At any rate, a clear signal of the resonance is found. Obviously, our calculations do not show any resonant signal around $3960\,\MeV$, since a state at that energy has not been included in our amplitudes. The line shapes shown here could therefore serve as a baseline of future theoretical or experimental analysis to detect or discard the presence of an eventual additional state around $3960\,\MeV$.

\begin{figure}
\centering
\includegraphics[width=0.47\textwidth,keepaspectratio]{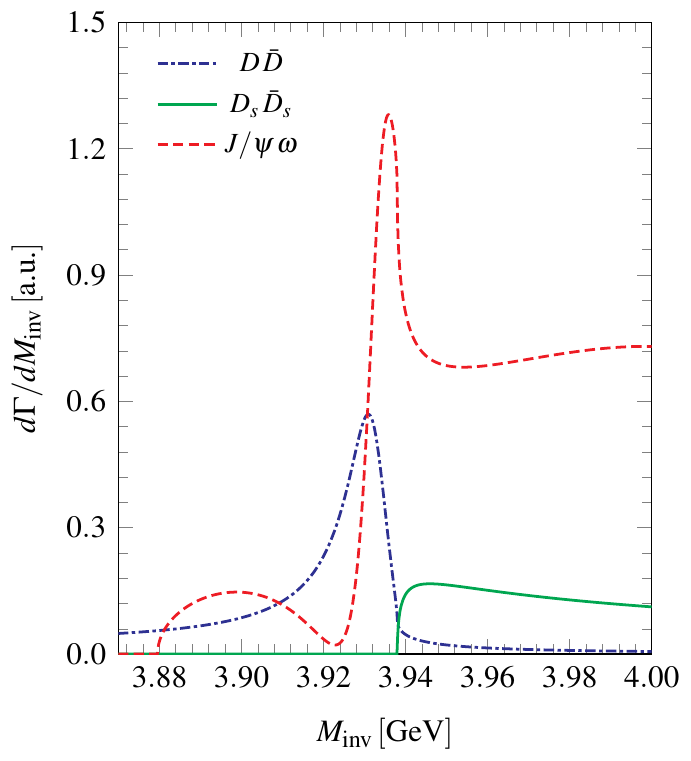}
\caption{Mass distributions $ d \Gamma / d M_\text{inv}(J/\psi \omega) , d \Gamma / d M_\text{inv}(D^+ D^- )  $  and $ d \Gamma / d M_\text{inv}( D_s ^+ D_s ^-)  $ for the reactions $B^+ \to K^+ J/\psi \,\omega $, $B^+ \to K^+ D^+ D^- $  and $B^+ \to K^+ D_s ^+ D_s ^- $, respectively. The last two mass distributions have been extracted from Ref.~\cite{Bayar:2022dqa}. }
\label{fig9}
\end{figure}

In Fig.~\ref{fig9} we compare the calculated mass distribution $ d \Gamma / d M_\text{inv}(J/\psi \omega)$  taking the normalization $ (C =1, C^{\prime}=-1, C^{\prime \prime}=3)$, with those studied in Ref.~\cite{Bayar:2022dqa}, considering $ M_\text{inv}(D^+ D^- ), M_\text{inv}( D_s ^+ D_s ^-)  $ for the reactions  $B^+ \to K^+ D^+ D^- $  and $B^+ \to K^+ D_s ^+ D_s ^- $. The scattering amplitudes for these two reactions correspond to the mechanism of Fig.~\ref{fig3} removing the triangle loop and is then given by (see also~\cite{Bayar:2022dqa})
\begin{eqnarray}
\tilde{t}_{D^+ D^-} & = & C^{\prime} \, G_{ D \overline{D}}(M_{\rm inv})\, T_{ D \overline{D}, D \overline{D}} (M_{\rm inv}) ,
 \nonumber \\
\tilde{t}_{D_s ^+ D_s ^-}  & = &  \frac{C^{\prime}}{\sqrt{2}} \,G_{ D_s ^+ D_s ^- }(M_{\rm inv}) \,T_{ D_s ^+ D_s ^-, D \overline{D}} (M_{\rm inv})  .
             \label{eq29}
\end{eqnarray}
From Fig.~\ref{fig9} we conclude that the expected signal is of the same order of that observed for $B^- \to K^- D^+ D^- $ and is reachable with the statistics of present runs and should be even clearer in future runs of LHCb.

\begin{figure}
\centering
\includegraphics[width=0.47\textwidth,keepaspectratio]{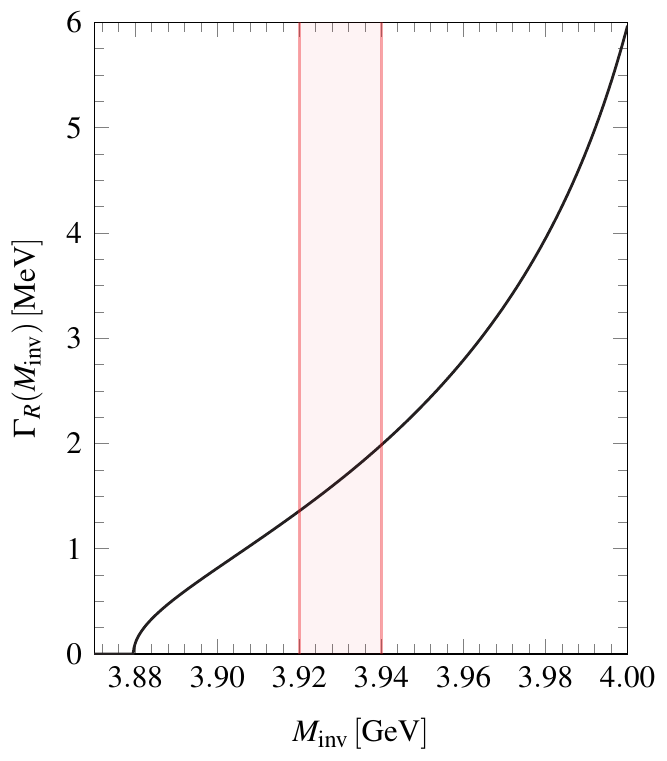}
\caption{Contribution to the width $ \Gamma_{X_0(3930)} $ from the $J/\psi \omega $ channel, as a function of the $ J/\psi \omega $ invariant mass. The vertical lines delimit the region of the resonance. }
\label{fig10}
\end{figure}

Finally, in Fig.~\ref{fig10} we plot the width $ \Gamma_{X_0(3930)} $ from the $J/\psi \, \omega $ decay channel given by Eq.~(\ref{eq23}), as a function of the $ J/\psi \, \omega $ invariant mass implicit in the  $ \widetilde{V} $ function, defined in the coupling $ g_ {X_0(3930) J/\psi \omega} $. Taking the mass of the resonance considered here, one can conclude that the contribution to the width $ \Gamma_{X_0(3930)} $ from the $J/\psi \omega $ channel is around $ 1.7 \,\MeV$. 
This value is about one order of magnitude smaller than the width coming from the decay to $ D \overline{D}$. Even if the coupling of the resonance to $ D \overline{D}$ is small compared to that to $ D_s ^+ D_s ^- $, the width for the decay into the $ D \overline{D}$ channel is still larger than the one into $ J/\psi \, \omega $. Then it might be surprising to see that the signals in Fig.~\ref{fig8} are of comparable order. This is due to the fact that the signal for  $ J/\psi \, \omega $ comes from the interference with a large tree level, and hence is linear, not quadratic in the resonance amplitude. 

We have checked uncertainties coming from changes in $ q_\text{max} $, by evaluating the results  with the use of $ q_\text{max} = 700\,\MeV$ and $ q_\text{max} = 800\,\MeV$. The changes in the results of Fig.~\ref{fig8} are small, at the level of $ 7\% $ and do not change qualitatively what is found there. The modifications in the $ X_0(3930) $ width for $ J/\psi \omega $ are larger, because it is quadratic in the resonant amplitude. On the other hand, the results for $ d \Gamma / d M_\text{inv} $ in Fig.~\ref{fig8} are linearly dependent on the resonance signal. Numerically we find 
\begin{eqnarray}
\Gamma_{X_0(3930)} (J/\psi \omega) & \approx & 
\left\{ \begin{array}{lcr}
1.51\,\MeV & \text{ for } & q_\text{max} = 700\,\MeV~, \\
1.72\,\MeV & \text{ for } & q_\text{max} = 750\,\MeV~, \\
2.04\,\MeV & \text{ for } & q_\text{max} = 800\,\MeV~.
\end{array} \right.
\label{eq30}
\end{eqnarray}
Hence, the changes are around $ 15 \% $. Yet, we still find that this width is reasonably smaller than the one for $ X_0(3930) $ decay to $ D \overline{D}$ .

The data in~Refs.~\cite{Andreassi:2014skr,BaBar:2010wfc} are indicative and promising. A clear peak is seen in the $ J/\psi \omega$ mass distribution in the $B^+ \to K^+ J/\psi \omega $ decay. It is unclear, however, how much of the peak can be due to a $ J/\psi \omega$ in $S$-wave with $J=0$ or in $D$-wave with $J=2$. This is actually the case in the $D^+ D^-$ mass distribution in the LHCb experiment~\cite{LHCb:2020bls,LHCb:2020pxc}, where the peak in this region is found to come from the $X_0(3930)$ and $\chi_{c2}(3930)$.  Disentangling  the content of the peak along the lines used in the LHCb analysis~\cite{LHCb:2020bls,LHCb:2020pxc} is necessary to allow a comparison with the predictions in the present work. In any case, it is already illustrative to see that in this experiment one cannot find a trace of a possible peak around $3960$~MeV. This feature is even clearer in the $\gamma^* \gamma^* \to  J/\psi \omega$ spectrum of~Ref.~\cite{BaBar:2012nxg}. These experiments, so far, do not support the hypothesis that there is an extra state around $3960\,\MeV$.

\section{Conclusions}

We have studied the $B^- \to K^- J/\psi \omega $ decay, paying attention to the formation of the state $X_0(3930)$, which in our approach comes from the interaction of the coupled channels $D \overline{D}$ and $D^{+}_s D^{-}_s$. Prior to the present work we had looked at the $D^+ D^-$ and $D^{+}_s D^{-}_s$ mass distributions in the $B^- \to K^- D^+ D^-$ and $B^- \to K^- D_s^+ D_s^-$ reactions, reaching the conclusion that the $X(3960)$ state, claimed from the peak around the $D_s^+ D_s^-$ threshold in the $B^- \to K^- D_s^+ D_s^-$ reaction was actually the same state $X_0(3930)$ observed in the $D^+ D^-$ mass distribution in the $B^- \to K^- D^+ D^-$ decay. Indeed, the existence of a resonance below the threshold of some channel induces an enhanced mass distribution around the threshold of that channel if the state couples to it. Such a state below the  $D^{+}_s D^{-}_s$ threshold is found in Refs.~\cite{Bayar:2022dqa,Ji:2022uie,Ji:2022vdj} and also in the lattice results of Ref.~\cite{Prelovsek:2020eiw}. The purpose of the present work is to offer additional information to corroborate, or refute, this reasonable hypothesis. For this purpose we have used the same formalism and input as in Ref.~\cite{Bayar:2022dqa} and implemented the transition from the $D \overline{D}$ and $D_s^+ D_s^-$ components to $J/\psi \omega $ via a triangle loop diagram with $D$ exchange. In addition, we have also considered the unavoidable tree level contribution to $B^- \to K^- J/\psi \omega $, which we have added to the resonant mechanisms. 
   
  We have some unknown magnitudes in the approach, but, up to a normalization not relevant in the study, we control the order of magnitude of these magnitudes and we study different possibilities within the freedom that we have. We find important consequences of our study. The first one is that the tree level and the resonant terms add coherently and we find clear interference of the amplitudes, constructive some times and destructive other times, but the signal of the resonance is clearly seen either way. This should serve as a warning for experimental analyses, where commonly the non resonant part, background, is added incoherently to the resonant contributions. The other relevant consequence is that, since the threshold of  $J/\psi \omega $ is at $3880\,\MeV$, far below the $D_s^+ D_s^-$ threshold, then the decay to this channel is open for $3930\,\MeV$ and $3960\,\MeV$. With our hypothesis that there is only one state at $3930\,\MeV$, we obviously observe only the resonant signal around this energy, but the shape of this signal, whether constructive or destructive, can be approximately controlled and, furthermore, is relatively narrow. Should there be an extra state at around $3960\,\MeV$ or, more generally, above the $D_s^+ D_s^-$ threshold, we expect to find an additional peak at that energy. Hence, the performance of the experiment will tell us whether there is one or two states.

   So far there are data on this reaction from the Thesis work of Guido Andreassi \cite{Andreassi:2014skr} in 2014. However, the data are not yet published. The statistics of the data do not allow us to make conclusions at present, but, with increased number of data in present and future runs of LHCb, the spectra of this reaction should be precise enough to provide an answer to the questions raised in the present work.

As for the BaBar analysis of~Ref.~\cite{BaBar:2010wfc}, where a peak is clearly seen in the $ J/\psi \omega $ mass distribution of the $B^+ \to K^+ J/\psi \omega $ decay, as well as for the work of~Ref.~\cite{Andreassi:2014skr}, it is also important to perform a separation of the contribution of $J=0 $ and $J=2$, once it has been shown that the peak at this energy seen in the $D^+ D^-$ mass distribution in the LHCb experiment in $B^+ \to K^+ D^+ D^- $ decay comes from the contribution of the $X_0(3930)$ and the $\chi_{c2}(3930)$. The techniques to do this separation are, thus, available and their use in further analysis of the $B^+ \to K^+ J/\psi \omega $ reaction is most welcome. Moreover, from the data of these works, and also from $\gamma^* \gamma^* \to  J/\psi \omega$~\cite{BaBar:2012nxg}, one cannot find any signature of the existence of a state at $3960$~MeV.

\begin{acknowledgements}
Discussions with M.\,Mikhasenko are much appreciated. This work was supported by the Spanish Ministerio de Ciencia e Innovaci\'on (MICINN) and European FEDER funds under Contracts No.\,PID2020-112777GB-I00, and by Generalitat Valenciana under contract PROMETEO/2020/023. This project has received funding from the European Union Horizon 2020 research and innovation programme under the program H2020-INFRAIA-2018-1, grant agreement No.\,824093 of the STRONG-2020 project. M.\,A. is supported through Generalitat Valencia (GVA) Grant No.\,CIDEGENT/2020/002. The work of A.\,F. was partially supported by the Generalitat Valenciana and European Social Fund APOSTD-2021-112. L.\,M.\,A. has received funding from the Brazilian agencies Conselho Nacional de Desenvolvimento Cient\'ifico e Tecnol\'ogico (CNPq) under contracts 309950/2020-1, 400215/2022-5, 200567/2022-5), and Funda\c{c}\~ao de Amparo \`a Pesquisa do Estado da Bahia (FAPESB) under the contract INT0007/2016. M.\,A. and A.\,F. thank the warm support of ACVJLI.
\end{acknowledgements}

\appendix\label{app:prueba}

\section{Triangle loop reduction to two-point loop function}\label{app:integral}

\newcommand{\ie}{i\epsilon}
\newcommand{\fL}{f_\Lambda}

In Eq.~\eqref{eq1415} we end up with the following loop integral:
\begin{equation}
I_\Lambda(s) = \int \frac{d^3 \vec{q}}{(2\pi)^3} \frac{1}{\omega_D(q)} \frac{\vec{q}^{\,2} \, \fL(\vec{q}^{\,2})}{s - 4\omega^2_D(q) + \ie}~,
\end{equation}
where $\fL(\vec{q}^{\,2}) = \thetaCO \left( q_\text{max} - \left\lvert \vec{q} \right\rvert \right)$ is our sharp cutoff regulator, although its specific form will only be used in the final step. This integral is equivalent to a two-point loop integral, but with a $\vec{q}^{\,2}$ factor in the numerator, so we can try and write it in terms of $G_{\Lambda}(s)$, the two-point scalar loop function\footnote{Similar results would be obtained here if we followed the Passarino-Veltman method of Refs.~\cite{tHooft:1978jhc,Passarino:1978jh}, see \textit{e.g.} Refs.~\cite{AlbaladejoSerrano:2012mua,Albaladejo:2012te}.} regularized with the regulator $\fL(\vec{q}^{\,2})$. We write the on-shell momentum as $p_D^2(s) = s/4 - m_D^2$, and the denominator can be written as:
\begin{equation}
D(s,q^2) = s - 4\omega^2_D(q) + \ie = -4\left( q^2 - p_D^2(s) \right) + \ie~.
\end{equation}
We subtract and add the on-shell momentum in the numerator, and we find:
\begin{align}
I_\Lambda(s) & = \displaystyle \int \frac{d^3 \vec{q}}{(2\pi)^3} \frac{\fL(\vec{q}^{\,2})}{\omega_D(q)} \, \frac{\vec{q}^{\,2} - p_D^2(s) + p_D^2(s)}{-4\left( q^2 - p_D^2(s) \right) + \ie} \nonumber\\
     & = \displaystyle  -\frac{1}{4} \int \frac{d^3 \vec{q}}{(2\pi)^3} \frac{\fL(\vec{q}^{\,2})}{\omega_D(q)}   + p_D^2(s)      \int \frac{d^3 \vec{q}}{(2\pi)^3} \frac{1}{\omega_D(q)} \frac{\fL(\vec{q}^{\,2})}{s - 4\omega^2_D(q) + \ie} \equiv \displaystyle -\frac{1}{4} \Omega_\Lambda + p_D^2(s) G_\Lambda(s)~,
\end{align}
being $\Omega_\Lambda$ just a number. Taking a sharp cutoff regulator, as pointed out above, we have indeed:
\begin{equation}
(2\pi)^2\,\Omega_\Lambda = \Lambda \sqrt{\Lambda^2 + m_D^2} + \frac{m_D^2}{2} \log \frac{\sqrt{\Lambda^2 + m_D^2} - \Lambda}{\sqrt{\Lambda^2 + m_D^2} + \Lambda}~.
\end{equation}
In addition, an explicit, algebraic expression for $G_\Lambda(s)$ can be found in (the erratum to) Ref.~\cite{Oller:1998hw}. Physically, this reduction of a three- to a two-point loop function means that, because the exchanged $D$-meson is far off-shell and its propagator is factored out, we have reduced the three-point loop integral with a $D$-meson exchange to a two-point loop function and a contact interaction for the amplitude $D\overline{D} \to J\psi\,\omega$.



\bibliographystyle{JHEP}
\bibliography{J_psi_omega.bib}   

\end{document}